\documentclass[11pt, reqno]{amsart}
\usepackage{hyperref}

\begin{document}
\title[Pricing Corporate Defaultable Bond ...]
{ Pricing Corporate Defaultable Bond using Declared Firm Value}

\author[Hyong-Chol O, Jong-Jun Jo and Chol-Ho Kim]{Hyong-Chol O, Jong-Jun Jo and Chol-Ho Kim}

\address{Hyong-Chol O \newline
Faculty of Mathematics,  Kim Il Sung University,  Pyongyang, D. P. R. Korea}
\email{ohyongchol@yahoo.com}

\address{Jong-Jun Jo   \newline
Faculty of Mathematics,  Kim Il Sung University,  Pyongyang, D. P. R. Korea}

\address{Chol-Ho Kim  \newline
Faculty of Mathematics,  Kim Il Sung University,  Pyongyang, D. P. R. Korea}

\thanks{First version submitted: on 15 Feb 2013, last revised: on 1 July 2013.}
\subjclass[2010]{35C15, 35Q91, 91G20, 91G40, 91G50, 91G80}
\keywords{defaultable bond, expected default, unexpected default, default intensity, partial differential equation}

\setcounter{page}{1}

\maketitle
\begin{abstract}
We study the pricing problem for corporate defaultable bond from the viewpoint of the investors outside the firm that could not exactly know about the information of the firm. We consider the problem for pricing of corporate defaultable bond in the case when the firm value is only declared in some fixed discrete time and unexpected default intensity is determined by the declared firm value. Here we provide a partial differential equation model for such a defaultable bond and give its pricing formula. Our pricing model is derived to solving problems of partial differential equations with random constants (default intensity) and terminal values of binary types. Our main method is to use the solving method of a partial differential equation with a random constant in every subinterval and to take expectation to remove the random constants. 
\end{abstract}

\section{Introduction}
There are two main approaches to pricing defaultable corporate bonds; one is the {\it structural approach} and the other one is the {\it reduced form approach}.\\\indent
In the structural method, we think that the default event occurs when the firm value is not enough to repay debt, that is, when the firm value reaches a certain lower threshold (called {\it default barrier}) from the above. Such a default can be expected and thus we call it {\it expected default}.\\\indent
In the reduced-form approach, the default is treated as an unpredictable event governed by default intensity process. In this case, the default event can occur without any correlation with the firm value and such a default is called {\it unexpected default}. In the reduced-form approach, if the default probability in time interval $[t, t+\Delta t]$ is $\lambda \Delta t$, then $\lambda$ is called a {\it default intensity}. \\\indent
If an investor knows all information about the firm value and default barrier in every time, then it is better for him to use the structural approach. If an investor can not exactly know about the firm value or default barrier, then he needs to use reduced form model.\\\indent
Nowadays, the use of unified models of structural approach and reduced-form approach is a trend \cite{BaP, BiB, CaE1, CaE2, OW, rea}. For example, using unified models of structural approach and reduced-form approach, Realdon \cite{rea} studied a pricing of corporate bonds in the case with constant default intensity and gave pricing formulae of the bond using PDE method. Cathcart et al. \cite{CaE1} studied a pricing of corporate bonds in the case when the default intensity is a linear function of the interest rate. They gave a semi-analytical pricing formula. Cathcart et al.\cite{CaE2} presented a valuation model that combines features of both the structural and reduced-form approaches for modeling default risk. In \cite{CaE2} the default intensity is a linear function of the state variable and the interest rate and they found that term structures of credit spreads generated using the middle-way approach were more in line with empirical observations. Other authors studied the pricing model of defaultable bonds in which the default intensity is given as a stochastic process \cite{BaP, BiB, OW, rea, wil}.  In \cite{OW}, the authors provided analytical pricing formula of corporate defaultable bond with both expected and unexpected defaults in the case when stochastic default intensity follows one of 3 special cases of Wilmott model \cite{wil}. Bi et al. \cite{BiB} provided the similar result with \cite{OW} in the case when stochastic default intensity follows CIR-like model. Ballestra et al. \cite{BaP} proposed a new model to price defaultable bonds which incorporates features of both structural and reduced-form models of credit risk where default intensity is described by an additional stochastic differential equation coupled with the process of the firm's asset value, and provided a closed-form approximate solution to their model.\\\indent
In the papers \cite{BaP, BiB, CaE1, CaE2, OW, rea}, they tried to express the price of the bond in terms of the firm value or the related signal variable to the firm value and the values of default intensity and defaul barrier at any time in the whole lifetime of the bond. \\\indent
On the other hand, every company announces its management data once in a certain term (for example, every quarter or every six months) and the announced data reflect the firm's financial circumstances. It is difficult for investors outside of the firm to know the firm's financial data except for these discrete announcing dates. \\\indent
According to this circumstance, in this paper we study the pricing problem for defaultable corporate bond from the viewpoint of the investors outside the firm that could only know the time-discretely announced information of the firm. We assume that we only know the firm value and the default barrier at several fixed discrete announcing dates and we don’t know about any information of the firm value in another time. We assume that the default intensity between the adjoined two announcing dates is determined by its announced firm value at the former announcing date and it is not changed in that time interval. And we assume that the firm value follows a geometric Brownian motion. (This problem was studied in \cite{OWR} but it included an error in deriving the pricing formula.) Such an approach is a kind of study of defaultable bond under {\it insufficient information} about the firm and it is interesting to note that Agliardi et al. \cite{AA} studied bond pricing problem under {\em imprecise information} with the technique of fuzzy mathematics.\\\indent
In this paper, when pricing corporate defaultable bond, we use the reduced form approach on every time interval between the adjoined two announcing dates and use the structural approach at the announcing dates. And although we take the unexpected default intensity as a constant, but we assume that the unexpected default intensity between the adjoined two announcing dates depends on its announced firm value at the former announcing date. Thus we try to use all available information we can get. We suppose such an approach would comparatively be reasonable.\\\indent
Characteristics of our model are 1) the starting point is the viewpoint of investors outside of the firm that could not exactly know the firm value and default information; 2) our model is one of structural-reduced form unified model. \\\indent
In our model, the short rate follows a generalized Hull-White model. The default event occurs in expected manner when the firm value reaches a certain lower threshold - the default barrier at one of the announcing dates or in unexpected manner at the first jump time of a Poisson process with intensity, respectively. Then our pricing problem is derived to a solving problem of PDE with random constant default intensity and terminal value of binary type in every subinterval between the two adjoined announcing dates. \\\indent
Our main method to solve this problem is to use the solving method of a partial differential equation with a random constant in every subinterval between the two adjoined announcing dates and to take expectation to remove the random constant.\\\indent
The remainder of the article is organized as follows: in section 2 we provide our modeling on corporate bond problem and give the pricing formula. In section 3 we prove the pricing formula.
%
\section{Modeling and the Pricing Formula}
{\bf Assumptions}\\\indent    
1) A firm issues a corporate bond with maturity $T$ and maturity face value 1.\\\indent
2) Let $0=t_0<t_1<\cdots<t_N-1<t_N=T$. At every time $t_i$, the firm value $V_i=V(t_i)$ is revaluated and announced. The firm value  $V(t)$ follows a geometric Brown motion 
\begin{equation*}
dV(t)=(\mu -b)V(t)dt+s_{V}V(t)dW_{1}(t)
\end{equation*}
under the risk neutral martingale measure. ($\mu,~b$ and $s_V$ are constants and $W_1(t)$ is an 1-dimensional standard Wiener process.) The firm continuously pays out dividend in rate $b$ for a unit of firm value. \\\indent
3) The unexpected default probability in the interval $[t, t+\Delta t]\cap [t_i, t_{i+1})$ is $\lambda_{i}\Delta t$, and the default intensity $\lambda_i$ is a known deterministic function of the firm value $V_i$ at the time $t_i$. For example, if we can assume that $\lambda_i=\lambda (V_i)=\ln(1+\frac{1}{V_i})$, then $\lambda_i$ goes to 0 when $V$ goes to infinity. This can be compatible with the real situations.\\\indent
4) Short rate satisfies the following condition under the risk neutral martingale measure:
\begin{equation}
dr_t=a_{r}(r, t)dt+s_{r}(t)dW_{2}(t),~~~a_{r}(r, t)=a_1(t)-a_2(t)r. \label{1}
\end{equation}
Here $W_2(t)$ is an 1-dimensional standard Wiener process.\\\indent
5) The unexpected default recovery $R_{ud}$ is given by $R_{u}\cdot Z(r, t)$ (exogenous recovery). Here recovery rate $0\leq R_{u}\leq 1$ is a constant and $Z(r, t)$ is the price of the default free zero coupon bond.\\\indent
6) The expected default barrier is only given at the time $t_i$. Expected default event occurs when $V(t_i)\leq K_i$. Here $K_i$ is a constant and the expected default recovery $R_{ed}$ is given by $R_{e}\cdot Z(r, t)$, where recovery rate $0\leq R_{e}\leq 1$ is a constant. \\\\\indent
{\bf Method of Modeling} \\\indent   
For simplicity, we assume that $N=2$. From assumption 3), after the time $t_1$ the unexpected default intensity $\lambda_1=\lambda(V_1)$ in the subinterval $(t_1, T]$ is a {\em known constant} and there is no any expected default in the open interval $(t_1,T)$. And at the time $t=T$ expected default event occurs when $V_2<K_2$. So under the condition that the firm value $V_1$ at the time $t_1$ is already known, the bond price $C_1(r, t;V_2|V_1)$ in the interval $(t_1,T]$ (when we regard the firm value $V_2$ at the time $t_2=T$ as a known quantity) can be seen as a derivative of short rate with the constant default intensity $\lambda(V_1)$ and it satisfies the following reduced-form model \cite{wil}:
\begin{eqnarray}
\frac{\partial C_1}{\partial t}+\frac{s_r^{2}(t)}{2}\frac{\partial^2 C_1}{\partial r^2}+a_r(r,t)\frac{\partial C_1}{\partial r}-rC_1+\lambda(V_1)(R_uZ(r,t)-C_1)=0,             \label{2} \\   
C_1(r,t_2)=C_1(r,t_2;V_2|V_1)=\left\{
\begin{array}{rl}
1~~ & \text{if} ~~~~~V_2>K_2,\\ 
R_e & \text{if} ~~~~~V_2\leq K_2.                               \label{3}      
\end{array} \right.
\end{eqnarray}
Here $Z(r,t)$  is default free zero coupon bond price and $V_2$  at the time $t<T$ is, in fact, an unknown random parameter. We solve the problem \eqref{2}, \eqref{3} to get the function $C_1(r,t;V_2|V_1)$.\\\indent
From the assumption 2) we can get the distribution of $V_2$ under the condition that $V_1$ is known, and thus taking expectation in $C_1(r,t;V_2|V_1)$ on $V_2$ we can get our bond price $C(r,t;V_1)$ in the interval $(t_1,T]$.\\\indent
In the interval $[0, t_1]$ the unexpected default intensity $\lambda(V_0)$ is a known constant and at the time $t=t_1$ expected default event occurs when $V_1<K_1$. So for every fixed firm value $V_1$ (at the time $t_1$) the bond price $C_0(r,t;V_1|V_0)$ in the interval $[0,t_1]$ satisfies the following reduced-form model:
\begin{eqnarray}
\frac{\partial C_0}{\partial t}+\frac{s_r^{2}(t)}{2}\frac{\partial^2 C_0}{\partial r^2}+a_r(r,t)\frac{\partial C_0}{\partial r}-rC_0+\lambda(V_0)(R_uZ(r,t)-C_0)=0,             \label{4} \\  
C_0(r,t_1)=C_0(r,t_1;V_1|V_0)=\left\{
\begin{array}{rl}
C(r,t_1;V_1) & \text{if} ~~~~~V_1>K_1,\\ 
R_eZ(r,t_1) & \text{if} ~~~~~V_1\leq K_1.    \label{5}                      
\end{array} \right.
\end{eqnarray}
Here $V_1$  at the time $t<t_1$ is in fact an unknown random parameter, too.\\\indent
If we solve the problem \eqref{4}, \eqref{5} to get the function $C_0(r,t;V_1|V_0)$ and take expectation on $V_1$, then we can get our bond price $C(r,t;V_0)$ in the interval $[0,t_1)$.\\\indent
$V_2$ in the problem \eqref{4}, \eqref{5} and $V_1$ in the problem \eqref{2}, \eqref{3} are random constants independent on the variables $r$ and $t$ of our equation. Thus the problem \eqref{2}, \eqref{3} and the problem \eqref{4}, \eqref{5} are terminal problems of partial differential equations with random parameters. And the terminal value conditions \eqref{3} and \eqref{5} are the functions of binary type that alternatively take two values on conditions.\\\indent
The bond price $C(r,t;V_0)$ in the interval $[0,t_1)$ depends on not only the short rate $r$ and $t$ but also the initial firm value $V_0$ (at $t=0$) and default barriers $K_1, K_2$.\\\\\indent
{\bf The Pricing Formula}\\\indent    
We have the following pricing formula in the time interval $[0,t_1)$:\\\\\indent
$C(r,t;V_0,K_1,K_2)=$
\begin{align}
&=Z(r,t)\{e^{-\lambda(V_0)(t_1-t)}[R_{u}N_2(\alpha_1,\alpha_2:A)+R_{u}R_{e}N_2(\alpha_1,-\alpha_2:\tilde A)+I_{22}+I_{24}] \nonumber\\
&+[1-e^{-\lambda(V_0)(t_1-t)}]R_{u}N(\alpha_1)+[R_{u}+(1-R_{u})e^{-\lambda(V_0)(t_1-t)}]R_{e}N(-\alpha_1)\}. \label{6}                  
\end{align}
Here $Z(r, t)$ is the price of risk free bond given in the next section and
\begin{eqnarray*}
&&N_2(a,b:A)=\frac{\sqrt{\det A}}{2 \pi}\int_{-\infty}^{a}\int_{-\infty}^{b}e^{-\frac{1}{2}\xi^{\bot}\cdot A\xi}dxdy,~~~\xi=(x,y)^{\bot},\\
&&\alpha_1=\frac{1}{s_V\sqrt{t_1}}\left[\ln\frac{V_0}{K_1}+(\mu-b-\frac{s_V^2}{2})t_1\right],\\
&&\alpha_2=\frac{1}{s_V\sqrt{t_2-t_1}}\left[\ln\frac{V_0}{K_2}+(\mu-b-\frac{s_V^2}{2})t_2\right],\\
&&A=\left(
\begin{array}{cc}
\frac{t_2}{t_2-t_1}  &  \sqrt{\frac{t_1}{t_2-t_1}}  \\
\sqrt{\frac{t_1}{t_2-t_1}}  &  1
\end{array}  \right),~ \tilde A=\left(
\begin{array}{cc}
\frac{t_2}{t_2-t_1}  &  -\sqrt{\frac{t_1}{t_2-t_1}}  \\
-\sqrt{\frac{t_1}{t_2-t_1}}  &  1
\end{array}  \right), \\
&&N(a)=\frac{1}{\sqrt{2 \pi}}\int_{-\infty}^{a}e^{-\frac{1}{2}x^2}dx,\\
&&I_{22}=(1-R_{u})\frac{1}{\sqrt{2 \pi}}\int_{-\infty}^{\alpha_1}F(x)N\left(\alpha_2+x\sqrt{\frac{t_1}{t_2-t_1}}
\right)e^{-\frac{x^2}{2}}dx,\\
&&I_{24}=(1-R_{u})R_e\frac{1}{\sqrt{2 \pi}}\int_{-\infty}^{\alpha_1}F(x)N\left(
-\alpha_2-x\sqrt{\frac{t_1}{t_2-t_1}}
\right)e^{-\frac{x^2}{2}}dx,
\end{eqnarray*}
Here
\[
F(x)=\exp\left\{-(t_2-t_1)\lambda\left(V_0e^{\left(\mu-b-\frac{s_V^2}{2}\right)t_1+s_Vx\sqrt{t_1}}\right)\right\},
\]
A proof of the formula \eqref{6} is provided in appendix.
\section{Appendix: Proof of the Pricing Formula}
Here we prove the formula \eqref{6}.\\\indent
Under the assumption 4) in the domain $K=\{(r, t)| r\in\textbf{R}, t\in[0,T]\}$, the price  $Z(r, t)$ of risk free bond satisfies the following problem:
\begin{equation}
\frac{\partial Z}{\partial t}+\frac{s_r^{2}(t)}{2}\frac{\partial^2 Z}{\partial r^2}+a_r(r,t)\frac{\partial Z}{\partial r}-rZ=0,~~Z(r,T)=1. \label{7}
\end{equation}
The solution is given by 
\begin{equation}
Z(r,t)=e^{A(t)-B(t)r},\label{8}
\end{equation}
Here $A(t)$ and $B(t)$ are differently given dependant on the specific models (including {\it Vasicek, Ho-Lee and Hull-White} models) of short rate \cite{wil}. For example, if the short rate satisfies the {\it Vasicek} model, that is, if the coefficients $a_1(t),~a_2(t), s_r(t)$ in \eqref{1} are all constants $a_1, a_2, s_r$, then $A(t)$ and $B(t)$ are given as follows \cite{wil}:
\[
B(t)=\frac{1-e^{-a_2(T-t)}}{a_2},~~A(t)=-\int_t^T[a_2B(u)-\frac{1}{2}s_r^2B^2(u)]du.
\]
\indent
{\bf Solving the problem \eqref{2} and \eqref{3}}\\\indent
In \eqref{2} and \eqref{3} we use the unknown function transformation $C_1(r,t)=u_1(t)Z(r,t)$ and consider the equation \eqref{7} and the relation \eqref {8}, then we have the following equation with $u_1(t)$ as an unknown function:
\begin{eqnarray*}
&&\frac{du_1}{dt}-\lambda(V_1)u_1+\lambda(V_1)R_u=0,~~(t_1<t<t_2=T), \\
&&u_1(T)=\left\{
\begin{array}{rl}
1~~ & \text{if} ~~~~~V_2>K_2,\\ 
R_e & \text{if} ~~~~~V_2\leq K_2.
\end{array} \right.
\end{eqnarray*}
It is an initial value problem of an ordinary differential equation and the solution is easily given by
\[
u_1(t)=\left\{
\begin{array}{rl}
R_u+(1-R_u)e^{-\lambda(V_1)(t_2-t)}~~ & \text{if} ~~~~~V_2>K_2,\\ 
R_u+(R_e-R_u)e^{-\lambda(V_1)(t_2-t)} & \text{if} ~~~~~V_2\leq K_2.
\end{array} \right.
\]
Thus the solution to \eqref{2} and \eqref{3} is given by
\begin{equation}
C_1(r,t;V_2|V_1)=\left\{
\begin{array}{rl}
Z(r,t)[R_u+(1-R_u)e^{-\lambda(V_1)(t_2-t)}]~~ & \text{if} ~~~~~V_2>K_2,\\ 
Z(r,t)[R_u+(R_e-R_u)e^{-\lambda(V_1)(t_2-t)}] & \text{if} ~~~~~V_2\leq K_2. \label{9}
\end{array} \right.
\end{equation}
\\\indent
{\bf The price of the Bond in the time interval $(t_1,T]$}\\\indent
From the assumption 2) we have 
\begin{eqnarray*}
&&V_t=V_s\exp\left[(\mu-b-\frac{s_V^2}{2})(t-s)+s_V(W_{1t}-W_{1s})\right],\\
&&\text{Prob}\{W_{1t}-W_{1s}\in A\}=\int_A\frac{1}{\sqrt{2\pi(t-s)}}\exp\left[-\frac{x^2}{2(t-s)}\right]dx.
\end{eqnarray*}
Thus we have  
\[\text{Prob}\{V_{2}>K_2\}=\int_{-\frac{1}{s_V}[\ln \frac{V_1}{K_2}+(\mu-b-\frac{s_V^2}{2})(t_2-t_1)]}^{\infty}\frac{1}{\sqrt{2\pi(t_2-t_1)}}\exp\left[-\frac{x^2}{2(t_2-t_1)}\right]dx\]
\begin{eqnarray*}
&&~=\int_{-\infty}^{\frac{1}{s_V\sqrt{t_2-t_1}}[\ln \frac{V_1}{K_2}+(\mu-b-\frac{s_V^2}{2})(t_2-t_1)]}\frac{1}{\sqrt{2\pi}}\exp\left[-\frac{x^2}{2}\right]dx.
\end{eqnarray*}
If in the above expression we use the cumulated distribution function $N(a)=\frac{1}{\sqrt{2 \pi}}\int_{-\infty}^{a}e^{-\frac{1}{2}x^2}dx$
of standard normal distribution and the notation of 
$$d_{-}(x/K,\mu,T-t)=\frac{\ln \frac{x}{K}+(\mu-b-\frac{s_V^2}{2})(T-t)}{s_V\sqrt{T-t}},$$
then we can get $\text{Prob}\{V_{2}>K_2\}=N[d_{-}(V_1/K_2,\mu,t_2-t_1)]$ and similarly we have
$\text{Prob}\{V_{2}\leq K_2\}=N[-d_{-}(V_1/K_2,\mu,t_2-t_1)].$ \\\indent 
We take expectation in \eqref{9} to remove the random constant $V_2$, then we have the price $C(r, t: V_1)$ of our bond in the interval $[t_1,T]$:
\begin{align*}
C(r, t: V_1)&=Z(r,t)\left[R_u+(1-R_u)e^{-\lambda(V_1)(t_2-t)}\right]\left\{N\left[d_{-}\left(\frac{V_1}{K_2},\mu,t_2-t_1\right)\right]+\right. \\
&+\left. R_eN\left[-d_{-}\left(\frac{V_1}{K_2},\mu,t_2-t_1\right)\right]\right\}  
\end{align*}
In particular, if we denote
\begin{align}
f(V_1)&=\left[R_u+(1-R_u)e^{-\lambda(V_1)(t_2-t_1)}\right]\left\{N\left[d_{-}\left(\frac{V_1}{K_2},\mu,t_2-t_1\right)\right]+\right. \nonumber \\
&+\left. R_eN\left[-d_{-}\left(\frac{V_1}{K_2},\mu,t_2-t_1\right)\right]\right\}  
\end{align}
then at the time $t_1$ we have
$$C(r, t_1: V_1)=Z(r,t_1)f(V_1).$$\indent
{\bf Solving of \eqref{4} and \eqref{5}}\\\indent
Now we know the price $C(r, t_1: V_1)$ of our bond at the time $t_1$ and thus the problem \eqref{4} and \eqref{5} on the interval $[0, t_1]$ is written as follows:
\begin{eqnarray}
\frac{\partial C_0}{\partial t}+\frac{s_r^{2}(t)}{2}\frac{\partial^2 C_0}{\partial r^2}+a_r(r,t)\frac{\partial C_0}{\partial r}-rC_0+\lambda(V_0)(R_uZ(r,t)-C_0)=0,   \label{11}\\      
C_0(r,t_1)=\left\{
\begin{array}{rl}
Z(r,t_1)f(V_1)~~ & \text{if} ~V_1>K_1,\\ 
R_eZ(r,t_1)\quad & \text{if} ~V_1\leq K_1. 
\end{array} \right. \label{12}                                                    
\end{eqnarray}
$V_0$ is known in the interval $[0, t_1]$, so $\lambda(V_0)$ is known constant. But $V_1$ is a random parameter in the interval $[0, t_1)$. For every fixed $V_1$, when $V_1\leq K_1$, we use the same method as the above to get the solution of \eqref{11} and \eqref{12}: 
\begin{equation}
C_0(r,t;V_1\leq K_1|V_0)=Z(r,t)[R_{u}+(R_e-R_u)e^{-\lambda(V_0)(t_1-t)}],~0\leq t<t_1.  \label{13}
\end{equation}
Similarly, when $V_1 > K_1$, we can get  
\begin{equation}
C_0(r,t;V_1> K_1|V_0)=Z(r,t)[R_{u}+(f(V_1)-R_u)e^{-\lambda(V_0)(t_1-t)}],~0\leq t<t_1. \label{14}
\end{equation}
\indent{\bf The price of the bond in time interval $[0, t_1]$}\\\indent
We add \eqref{13} and \eqref{14} after taking expectation on $V_1$ to remove it in \eqref{13} and \eqref{14}, then we have the price in the time interval $[0, t_1]$. As \eqref{13} does not include $V_1$ and we already knew 
$$\text{Prob}\{V_1\leq K_1\}=N[-d_{-}(V_0/K_1,\mu,t_1)],$$ 
as the above, we easily get the expectation of \eqref{13}:\\\\\indent 
$E(C_0(r,t;V_1\leq K_1|V_0))=$
\begin{equation}
=R_eZ(r,t)\left[R_u+(1-R_u)e^{-\lambda(V_0)(t_1-t)}\right]N\left[-d_{-}\left(\frac{V_0}{K_1},\mu,t_1\right)\right].                                            \label{15}
\end{equation}\indent
Now we calculate the expectation $E(C_0(r,t;V_1>K_1|V_0))$ of \eqref{14}. Unlike \eqref{13}, $C_0(r,t;V_1>K_1|V_0)$ is a function of $V_1$, and so we denote 
$$g(V_1)=g(r,t,V_0;V_1):=C_0(r,t;V_1> K_1|V_0).$$
From the assumption 2), 
\begin{equation}
V_1=V_0\exp\left[\left(\mu-b-\frac{s_V^2}{2}\right)t_1+s_VW_{1t_1}\right]       \label{16} 
\end{equation}
and thus $g(V_1)$ is written as
$$g(r,t,V_0;V_1)=g\left(r,t,V_0;V_0e^{\left(\mu-b-\frac{s_V^2}{2}\right)t_1+s_VW_{1t_1}}\right).$$
And we note that $\text{Prob}\{W_{1t}\in A\}=\int_A\frac{1}{\sqrt{2\pi t}}\exp\left[-\frac{x^2}{2t}\right]dx$, and $V_1>K_1 \Longleftrightarrow W_{1t}>-\frac{1}{s_V}\left[\ln \frac{V_0}{K_1}+\left(\mu-b-\frac{s_V^2}{2}\right)t_1 \right]$. Thus we have\\\\\indent  
$E(C_0(r,t;V_1>K_1|V_0))=$
\begin{equation}
~=\frac{1}{\sqrt{2\pi t_1}}\int_{-\infty}^{\frac{1}{s_V}\left[\ln \frac{V_0}{K_1}+(\mu-b-\frac{s_V^2}{2})t_1\right]}g\left(r,t,V_0;V_0e^{(\mu-b-\frac{s_V^2}{2})t_1+s_Vx}\right)e^{-\frac{x^2}{2t_1}}dx                \label{17}                             
\end{equation}
In order to calculate the integral of \eqref{17}, we need to get the representation of the function 
$$G(x)=g\left(r,t,V_0;V_0e^{(\mu-b-\frac{s_V^2}{2})t_1+s_Vx}\right)$$ 
in the integrand in \eqref{17}. From the definition of $g(r,t,V_0;V_1)$ and \eqref{14} we have
$$g(r,t,V_0;V_1)=R_{u}Z(r,t)\left[1-e^{-\lambda(V_0)(t_1-t)}\right]+Z(r,t)e^{-\lambda(V_0)(t_1-t)}f(V_1)$$
In \eqref{10}, if we write $f(V_1)$ as $f(V_1)=g_{21}(V_1)+g_{22}(V_1)+g_{23}(V_1)+g_{24}(V_1)$, then we have 
$$g(r,t,V_0;V_1)=g_1(r,t,V_0)+Z(r,t)e^{-\lambda(V_0)(t_1-t)}[g_{21}(V_1)+g_{22}(V_1)+g_{23}(V_1)+g_{24}(V_1)]$$
Here
\begin{eqnarray}
&&g_1(r,t,V_0)=R_{u}Z(r,t)\left[1-e^{-\lambda(V_0)(t_1-t)}\right], \nonumber\\
&&g_{21}(V_1)=R_uN\left[d_{-}\left(\frac{V_1}{K_2},\mu,t_2-t_1\right)\right], \nonumber\\ 
&&g_{22}(V_1)=(1-R_u)e^{-\lambda(V_1)(t_2-t_1)}N\left[d_{-}\left(\frac{V_1}{K_2},\mu,t_2-t_1\right)\right],                      \label{18} \\                          
&&g_{23}(V_1)=R_uR_eN\left[-d_{-}\left(\frac{V_1}{K_2},\mu,t_2-t_1\right)\right], \nonumber\\
&&g_{24}(V_1)=R_e(1-R_u)e^{-\lambda(V_1)(t_2-t_1)}N\left[-d_{-}\left(\frac{V_1}{K_2},\mu,t_2-t_1\right)\right]. \nonumber
\end{eqnarray}
In \eqref{16} we denote $x=W_{1t_1}$, then $V_1=V_0\exp\left[\left(\mu-b-\frac{s_V^2}{2}\right)t_1+s_Vx\right]$ and so we can write
$$e^{-\lambda(V_1)(t_2-t_1)}=\exp\left\{-(t_2-t_1)\lambda\left(V_0e^{(\mu-b-\frac{s_V^2}{2})t_1+s_Vx}\right) \right\}.$$ 
In particular, if the function $\lambda(V)$ is given by $\lambda(V)=\ln(1+\frac{1}{V})$, then we have 
\begin{equation}
e^{-\lambda(V_1)(t_2-t_1)}=\left[\frac{V_0e^{(\mu-b-\frac{s_V^2}{2})t_1+s_Vx}}{1+V_0e^{(\mu-b-\frac{s_V^2}{2})}t_1+s_Vx} \right]^{t_2-t_1}.    \label{19}
\end{equation}
Now we represent $d_{-}(V_1/K_2,\mu,t_2-t_1)$ in \eqref{18} as a function of $x$. If we denote
\begin{equation}
\alpha_2=\frac{\ln\frac{V_0}{K_2}+\left(\mu-b-\frac{s_V^2}{2}\right)t_2}{s_V\sqrt{t_2-t_1}},                        \label{20}
\end{equation}
then we have
$$d_{-}\left(\frac{V_1}{K_2},\mu,t_2-t_1\right)=\frac{\ln\frac{V_1}{K_2}+(\mu-b-\frac{s_V^2}{2})(t_2-t_1)}{s_V\sqrt{t_2-t_1}}=\alpha_2+\frac{x}{\sqrt{t_2-t_1}}.$$ 
Using this we get the representations of $g_{21},~g_{22},~g_{23}$ and $g_{24}$ in \eqref{18} in terms of $x$ (these are still writen as $g_{2i}$):
\begin{align}
&g_{21}(x)=R_u\frac{1}{\sqrt{2 \pi}}\int_{-\infty}^{\alpha_2}e^{-\frac{1}{2}\left(y+\frac{x}{\sqrt{t_2-t_1}}\right)^2}dy,                \label{21}\\ 
&g_{22}(x)=(1-R_u)\exp\left[-(t_2-t_1)\lambda(V_0e^{(\mu-b-\frac{s_V^2}{2})t_1+s_Vx})\right]N\left(\alpha_2+\frac{x}{\sqrt{t_2-t_1}}\right),    \label{22} \\
&g_{23}(x)=R_u\frac{1}{\sqrt{2 \pi}}\int_{-\infty}^{-\alpha_2}e^{-\frac{1}{2}\left(y-\frac{x}{\sqrt{t_2-t_1}}\right)^2}dy,     \label{23} \\
&g_{24}(x)=R_e(1-R_u)\exp\left[-(t_2-t_1)\lambda\left(V_0e^{(\mu-b-\frac{s_V^2}{2})t_1+s_Vx}\right)\right]N\left(-\alpha_2-\frac{x}{\sqrt{t_2-t_1}}\right). \label{24}
\end{align} 
Now we calculate \eqref{17}. \\\\  
$E(C_0(r,t;V_1>K_1|V_0))=$
\begin{align}
&=\frac{1}{\sqrt{2\pi t_1}}\int_{-\infty}^{\frac{1}{s_V}\left[\ln \frac{V_0}{K_1}+\left(\mu-b-\frac{s_V^2}{2}\right)t_1\right]}g\left(r,t,V_0;V_0e^{(\mu-b-\frac{s_V^2}{2})t_1+s_Vx}\right)e^{-\frac{x^2}{2t_1}}dx \nonumber  \\
&=\frac{1}{\sqrt{2\pi t_1}}\int_{-\infty}^{\frac{1}{s_V}\left[\ln \frac{V_0}{K_1}+\left(\mu-b-\frac{s_V^2}{2}\right)t_1\right]}g_1(r,t,V_0)e^{-\frac{x^2}{2t_1}}dx+  \nonumber \\
&+Z(r,t)e^{-\lambda(V_0)(t_1-t)}\frac{1}{\sqrt{2\pi t_1}}\int_{-\infty}^{\frac{\ln \frac{V_0}{K_1}+\left(\mu-b-\frac{s_V^2}{2}\right)t_1}{s_V}}f\left(V_0e^{(\mu-b-\frac{s_V^2}{2})t_1+s_Vx}\right)e^{-\frac{x^2}{2t_1}}dx \nonumber \\
&=I_1+Z(r,t)e^{-\lambda(V_0)(t_1-t)}I_2.            \label{25}
\end{align}
Here 
\begin{align}
I_1&=R_{u}Z(r,t)\left[1-e^{-\lambda(V_0)(t_1-t)}\right]\frac{1}{\sqrt{2\pi t_1}}\int_{-\infty}^{\frac{1}{s_V}\left[\ln \frac{V_0}{K_1}+\left(\mu-b-\frac{s_V^2}{2}\right)t_1\right]}e^{-\frac{x^2}{2t_1}}dx  \nonumber \\
&=R_{u}Z(r,t)\left[1-e^{-\lambda(V_0)(t_1-t)}\right]N\left[d_{-}\left(\frac{V_0}{K_1},\mu,t_1\right)\right],                          \label{26} \\
I_2&=\frac{1}{\sqrt{2\pi t_1}}\int_{-\infty}^{\frac{1}{s_V}\left[\ln \frac{V_0}{K_1}+\left(\mu-b-\frac{s_V^2}{2}\right)t_1\right]}f\left(V_0e^{(\mu-b-\frac{s_V^2}{2})t_1+s_Vx}\right)e^{-\frac{x^2}{2t_1}}dx \nonumber \\
&=\frac{1}{\sqrt{2\pi t_1}}\int_{-\infty}^{\frac{1}{s_V}\left[\ln \frac{V_0}{K_1}+\left(\mu-b-\frac{s_V^2}{2}\right)t_1\right]}\left[g_{21}(x)+g_{22}(x)+g_{23}(x)+g_{24}(x)\right]e^{-\frac{x^2}{2t_1}}dx \nonumber \\
&=I_{21}+I_{22}+I_{23}+I_{24}.                 \label{27}
\end{align}
Here we used the fact that $g_1(r,t,V_0)$ does not depend on $x$ and $f = g_{21}+g_{22}+g_{23}+g_{24}$.
Now we calculate $I_{2i}$. For simplicity of symbol, let denote 
\begin{equation}
\alpha_1=\frac{1}{s_V}\left[\ln \frac{V_0}{K_1}+\left(\mu-b-\frac{s_V^2}{2}\right)t_1\right] = d_{-}\left(\frac{V_0}{K_1},\mu,t_1\right).   \label{28}
\end{equation}
Then from \eqref{21}, we have
\begin{equation}
I_{21}=\frac{1}{\sqrt{2\pi}}\int_{-\infty}^{\alpha_1}g_{21}(x\sqrt{t_1})e^{-\frac{x^2}{2}}dx=\frac{R_u}{2\pi}\int_{-\infty}^{\alpha_1}dx\int_{-\infty}^{\alpha_2}e^{-\frac{x^2}{2}-\frac{1}{2}\left(y+\frac{x\sqrt{t_1}}{\sqrt{t_2-t_1}}\right)^2}dy.   \label{29}
\end{equation}
The exponent of the integrand of \eqref{29} can be written as a bivariate quadratic form $-\frac{1}{2}\xi^{\bot}A\xi$, where
\begin{equation}
A=\left(
\begin{array}{cc}
\frac{t_2}{t_2-t_1}  &  \sqrt{\frac{t_1}{t_2-t_1}}  \\
\sqrt{\frac{t_1}{t_2-t_1}}  &  1
\end{array}  \right),~\det A=1, \xi^{\bot}=(x,y).            \label{30}
\end{equation}
Thus $I_{21}$ is represented by the cumulated distribution function $N_2$ of the bivariate normal distribution as follows:
\begin{equation}
I_{21}=R_u\frac{\sqrt{\det A}}{2\pi}\int_{-\infty}^{\alpha_1}dx\int_{-\infty}^{\alpha_2}e^{-\frac{1}{2}\xi^{\bot}A\xi}dy=R_uN_2(\alpha_1,\alpha_2;A).   \label{31}
\end{equation}
Similarly, from \eqref{23}, we have the representation of $I_{23}$ by the cumulated distribution function $N_2$ of the bivariate normal distribution: 
\begin{equation}
I_{23}=R_uR_e\frac{\sqrt{\det \tilde A}}{2\pi}\int_{-\infty}^{\alpha_1}dx\int_{-\infty}^{-\alpha_2}e^{-\frac{1}{2}\xi^{\bot}\tilde A\xi}dy=R_uN_2(\alpha_1,-\alpha_2;\tilde A).     \label{32}
\end{equation}
Here
\begin{equation}
\tilde A=\left(
\begin{array}{cc}
\frac{t_2}{t_2-t_1}  &  -\sqrt{\frac{t_1}{t_2-t_1}}  \\
-\sqrt{\frac{t_1}{t_2-t_1}}  &  1
\end{array}  \right),~\det \tilde A=1.                     \label {33} 
\end{equation}
From \eqref{27} and \eqref{22} we directly get
\begin{align}
I_{22}&=\frac{1}{\sqrt{2\pi}}\int_{-\infty}^{\alpha_1}g_{22}(x\sqrt{t_1})e^{-\frac{x^2}{2}}dx=\nonumber \\
&=\frac{1-R_{u}}{\sqrt{2 \pi}}\int_{-\infty}^{\alpha_1}F(x)N\left(\alpha_2+x\sqrt{\frac{t_1}{t_2-t_1}}
\right)e^{-\frac{x^2}{2}}dx.                    \label{34} 
\end{align}
Here
\begin{align*}
F(x)=\exp\left\{-(t_2-t_1)\lambda\left(V_0e^{\left(\mu-b-\frac{s_V^2}{2}\right)t_1+s_Vx\sqrt{t_1}}\right)\right\}.     \label{35}
\end{align*}
From  \eqref{27} and \eqref{24} we directly get
\begin{align}
I_{24}=\frac{R_e(1-R_{u})}{\sqrt{2 \pi}}\int_{-\infty}^{\alpha_1}F(x)N\left(
-\alpha_2-x\sqrt{\frac{t_1}{t_2-t_1}}
\right)e^{-\frac{x^2}{2}}dx.
\end{align}
Substitute \eqref{31}, \eqref{32},\eqref{34} and \eqref{35} into \eqref{27} to get $I_2$. Then substitute $I_2$ and $I_1$ into \eqref{25} to get $E(C_0(r,t;V_1>K_1|V_0))$ of \eqref{17}. Then our bond price is given by
$$C(r,t;V_0)=E(C_0(r,t;V_1>K_1|V_0))+E(C_0(r,t;V_1\leq K_1|V_0))$$
which gives the above formula (6). \\\\\indent 
{\bf Acknowledgment} The authors would like to thank the managing editor Ahmed El-Sayed and Carlo Bianca for their help and advices for this article.

\end{document}